\documentclass[12pt]{article}

\usepackage{epsfig}
\usepackage{color,graphicx}
\usepackage{cite}
\usepackage{amssymb,amsmath}

\newcommand{\be}{\begin{equation}}
\newcommand{\ee}{\end{equation}}
\newcommand{\bea}{\begin{eqnarray}}
\newcommand{\eea}{\end{eqnarray}}

\begin{document}

\begin{flushright}
CCTP-2011-18 
\end{flushright}

\begin{center}  

\vskip 2cm 

\centerline{\Large {\bf Striped  instability of a holographic Fermi-like liquid}}
\vskip 1cm

\renewcommand{\thefootnote}{\fnsymbol{footnote}}

\centerline{Oren Bergman,${}^1$\footnote{bergman@physics.technion.ac.il} 
Niko Jokela,${}^{1,2}$\footnote{najokela@physics.technion.ac.il} 
Gilad Lifschytz,${}^2$\footnote{giladl@research.haifa.ac.il} 
and Matthew Lippert${}^3$\footnote{mlippert@physics.uoc.gr}}

\vskip .5cm
${}^1${\small \sl Department of Physics} \\
{\small \sl Technion, Haifa 32000, Israel} 

\vskip 0.2cm
${}^2${\small \sl Department of Mathematics and Physics} \\
{\small \sl University of Haifa at Oranim, Tivon 36006, Israel} 

\vskip 0.2cm
${}^3${\small \sl Crete Center for Theoretical Physics} \\
{\small \sl Department of Physics} \\
{\small \sl University of Crete,  71003 Heraklion, Greece}

\end{center}

\vskip 0.3 cm

\setcounter{footnote}{0}
\renewcommand{\thefootnote}{\arabic{footnote}}

\begin{abstract}
\noindent We consider a holographic description of a system of strongly-coupled fermions in 
$2+1$ dimensions based on a D7-brane probe in the background of D3-branes. 
The black hole embedding represents a Fermi-like liquid. We study the excitations of the Fermi liquid 
system. Above a critical density which depends on the temperature, the system 
becomes unstable towards an inhomogeneous modulated phase which is similar to a charge density and spin wave state. 
The essence of this instability can be effectively described by a Maxwell-axion theory with a background electric field.
We also consider the fate of zero sound at non-zero temperature.

\end{abstract}

\newpage

\section{Introduction}
In the last few years AdS/CFT has been used to analyze systems in which condensed-matter-like phenomena occur.
Among them are superconductivity and superfluidity (for a review see \cite{Hartnoll:2009sz}),
non-Fermi liquids \cite{Liu:2009dm,Cubrovic:2009ye}, metamagnetism \cite{Lifschytz:2009sz,D'Hoker:2010rz}, 
the quantum Hall effect \cite{Esko,Davis,Fujita,Hikida,Alanen,Bergman:2010gm,Jokela:2010nu,Jokela}, 
and more. In this note we provide another example, namely that of 
a striped phase of fermionic matter at non-zero density.

Some of the aforementioned phenomena have been investigated  in the bottom-up approach, 
in which one does not  start with a well-defined boundary theory, but rather with a toy model in the bulk which seems to 
capture the essence of the phenomena. 
Here we take the top-down point of view and study the D3-D7' model, a particular (2+1)-dimensional system consisting 
of a D7-brane intersecting a stack of D3-branes with $\#ND=6$.
We use the probe approximation, in which the D7-brane is treated as a probe in the near-horizon
background of the D3-branes.
The low-energy and weak-coupling spectrum has charged states that are purely fermionic.
It therefore
seems like a good place to start looking for interesting phenomena.

The D3-D7' system 
was originally proposed in \cite{Rey:2008zz} as a holographic model for strongly interacting fermions in 2+1 dimensions.
In \cite{Davis,Bergman:2010gm} this model was used  
to explore the phases of the system at non-zero temperature and charge density
in the background of a magnetic field. 
In \cite{Bergman:2010gm,Jokela:2010nu} it was shown that the system exhibits a gapped phase similar to a quantum
Hall state at fixed ratios of the magnetic field to the charge density.

Here we analyze the system in the ungapped phase and without a magnetic field. 
This system 
resembles a Fermi liquid in that the
resistivity at low temperature scales as $T^2$.
However, the heat capacity at low temperature does not follow the Fermi liquid formula. 

We 
analyze the four-dimensional bulk quasi-normal modes, which correspond to the poles in the bi-fermion two-point functions
of the three-dimensional boundary theory, including the current-current correlators. 
In particular, we will demonstrate that the system becomes unstable to the formation of a striped phase
when the ratio of the charge density to the squared temperature is above a critical value.
This instability can be effectively described by a $U(1)$ gauge field interacting with an axion field in a constant background electric field. 

Striped phases are known to exist in condensed matter systems. Examples include charge density waves and spin density waves 
(for reviews see \cite{Gruner:1994zz, Gruner:1988zz}), but there are more complicated situations as well. 
What is common to them is that some observable of the system is modulated in one direction and hence the name striped phase. 
Holographic striped phases in $3+1$ dimensions were analyzed in \cite{Nakamura:2009tf,Ooguri:2010kt,Ooguri:2010xs,Bayona:2011ab}.

\medskip

\noindent\underline{Note added}: As this paper was being completed  \cite{Donos:2011bh} appeared which finds a 
similar instability in a different (2+1)-dimensional holographic model.

\section{An instability in Maxwell-axion theory}

In \cite{Nakamura:2009tf} Nakamura, Ooguri, and Park demonstrated that five-dimensional
Maxwell-Chern-Simons theory becomes unstable in the presence of a background electric field
due to a non-zero momentum mode that becomes tachyonic.
As we will now show, a similar instability can occur in a four-dimensional Maxwell-axion theory, which has the Lagrangian density
\be
{\cal L} = -\frac{1}{4} F_{IJ}F^{IJ} + \frac{1}{2}(\partial_I \Phi)^2 - \frac{1}{2}m^2\Phi^2
+ \frac{\alpha}{2} \epsilon^{IJKL} \Phi F_{IJ}F_{KL} \,.
\ee
The equations of motion are given by 
\bea 
(\square + m^2) \Phi - 
\frac{1}{2}\alpha \epsilon^{IJKL} F_{IJ} F_{KL} &=& 0\\
\partial_I F^{IJ} - 2\alpha\epsilon^{IJKL}\partial_I(\Phi F_{KL}) &=& 0 \,.
\eea
Expanding around a background with $F=F^{(0)}$ and $\Phi = 0$,
the linearized equations for the fluctuations $f_{IJ}$ and $\varphi$ are given by
\bea 
(\square + m^2) \varphi - 
\alpha \epsilon^{IJKL} F^{(0)}_{IJ} f_{KL} &=& 0\\
\partial_I f^{IJ} - 2\alpha\epsilon^{IJKL}\partial_I(\varphi F^{(0)}_{KL}) &=& 0 \,.
\eea
Consider a constant background electric field in the $z$-direction $F^{(0)}_{03} = E$.
The equations for the fluctuations become
\bea 
(\partial_\mu \partial^\mu + \partial_i \partial^i + m^2)\varphi - 4\alpha E f_{12} &=& 0\\
\partial_\mu f^{\mu\nu} + \partial_i f^{i\nu} &=& 0 \\
\partial_\mu f^{\mu j} + \partial_i f^{ij} - 4\alpha E \epsilon^{ij}\partial_i \varphi  &=& 0 \,,
\eea
where $\mu,\nu$ take values in $(0,3)$ and $i,j$ take values in $(1,2)$.
Multiplying the third equation by $\epsilon_{jk}\partial^k$ and using the Bianchi identity gives
another equation involving only $f_{12}$ and $\varphi$,
\be
(\partial_\mu \partial^\mu + \partial_i \partial^i)f_{12} - 4\alpha E \partial_i \partial^i \varphi = 0 \,.
\ee
For plane wave solutions of the form 
\be
\varphi(x,t) = \bar{\varphi} e^{-i(\omega t - kx)} \; , \; f_{12}(x,t) = \bar{f}_{12}e^{-i(\omega t - kx)} \ ,
\ee
we obtain the dispersion relation
\be
 \omega^2=k^2+\frac{1}{2}m^2 \pm \frac{1}{2}\sqrt{m^4+64 \alpha^2 E^2 k^2} \ .
\ee
Therefore, there is a tachyonic mode in the range $0< k < \sqrt{16\alpha^2 E^2 - m^2}$.
For a background magnetic field, on the other hand, there is no instability.
We will find a similar instability in the four-dimensional bulk theory of the D3-D7' system.
This will be interpreted as an instability of the boundary theory at non-zero density towards the
formation of a striped phase.

\section{Review of the D3-D7' model}\label{sec:setup}

A holographic model for fermions in 2+1 dimensions can be constructed using
a D7-brane and D3-branes that intersect in 2+1 dimensions.
Since $\# ND=6$ in this case, the massless spectrum of the D3-D7 open strings 
consists only of the desired fermions.
The separation between the D7-brane and D3-branes in the transverse direction corresponds to the fermion mass.
We decouple the fermions from the closed string modes by focusing on the near-horizon limit of a large stack of $N_3$ D3-branes 
and treating the D7-brane as a probe.  
According to the usual gauge/gravity duality, the dynamics of the probe D7-brane in the near-horizon D3-brane 
background captures the physics of the (2+1)-dimensional fermions interacting with a strongly-coupled (3+1)-dimensional gauge theory.


\subsection{Background}

We begin with the near-horizon background of the non-extremal D3-branes:
\begin{eqnarray}
\label{D3metric}
 L^{-2} ds_{10}^2 &=& r^2 \left(-h(r)dt^{2}+dx^2+dy^2+dz^2\right)+
 r^{-2} \left(\frac{dr^2}{h(r)}+r^2 d\Omega_5^2\right) \\
\label{RR_5-form}
F_5 &=& 4L^4\left(r^3 dt\wedge dx\wedge dy\wedge dz\wedge dr 
+  d\Omega_5 \right) \,,
\end{eqnarray}
where $h(r)=1-r_T^4/r^4$ and $L^2=\sqrt{4\pi g_{s} N_3}\, \alpha'$. 
For convenience, we work in dimensionless coordinates, {\em e.g.}, $r=r_{phys}/L$.
This background is dual to ${\cal N}=4$ SYM theory at a temperature $T = r_T/(\pi L)$.
We parameterize the five-sphere as an $S^2\times S^2$ fibered over an interval:
\bea
 d\Omega_5^2 &=& d\psi^2 + 
 \cos^2\psi (d\Omega_2^{(1)})^2 + \sin^2\psi (d\Omega_2^{(2)})^2 \nonumber \\
 (d\Omega_2^{(i)})^2 &=& d\theta_i^2 + \sin^2\theta_i d\phi_i \,,
\eea
where $0\leq \psi \leq \pi/2$, $0\leq \theta_i \leq \pi$, and $0\leq \phi_i < 2\pi$.
As $\psi$ varies, the sizes of the two $S^2$'s change. At $\psi=0$ one of the $S^2$'s shrinks to zero
size, and at $\psi=\pi/2$ the other $S^2$ shrinks. The $S^2\times S^2$ at $\psi=\pi/4$ is the ``equator".

\subsection{Probe}

The D7-brane extends in $t,x,y,r$ and wraps the two two-spheres. 
The D7 embedding is usually taken to be  characterized by $\psi(r)$ and $z(r)$. However, excitations 
around this embedding are tachyonic. To cure this we turn on fluxes through the two two-spheres 
labeled by the parameters $f_1$ and $f_2$. With the correct choice of $f_1$ and $f_2$, one gets a stable embedding. 
We also consider a non-zero charge density, by including the time component of the worldvolume gauge field $a_0(r)$,
and a background magnetic field $b$.
The D7-brane action has a DBI term given by
\bea
\label{DBI_action}
 S_{DBI} & = & -T_7 \int d^8x\, e^{-\Phi} \sqrt{-\mbox{det}(g_{\mu\nu}+ 2\pi\alpha' F_{\mu\nu})} \nonumber \\
 &=& - {\cal N} \int dr\, r^2\sqrt{\left(4\cos^4\psi + f_1^2 \right)
 \left(4\sin^4\psi + f_2^2 \right)}\times \nonumber \\ 
 & & \qquad\qquad \times \sqrt{\left(1+ r^4 h z'^2+ r^2 h \psi'^2 - {a_0'}^2\right)\left(1+\frac{b^2}{r^4}\right)} \,,
\eea
and a CS term given by
\bea
\label{CS_action}
S_{CS} &=& -\frac{(2\pi\alpha')^2T_7}{2} \int P[C_4]\wedge F \wedge F \nonumber\\
&=&  -{\cal N}f_1 f_2 \int dr\, r^4 z'(r)
+ 2{\cal N} \int dr\, c(r) b a_0'(r) \,,
\eea
where ${\cal N} \equiv 4\pi^2 L^5 T_7 V_{2,1}$
and $c(r) = \psi(r) - \frac{1}{4}\sin\left(4\psi(r)\right) - \psi_\infty + \frac{1}{4}\sin(4\psi_\infty)$.\footnote{For the MN embedding that describes
the gapped phase, one must also add a boundary term $S_{boundary} =  2{\cal N} c(r_{min}) b a_0(r_{min})$.
However, we will only discuss BH embeddings, for which the boundary term vanishes.}
Note that $c(r)$, and therefore $\psi(r)$, plays the role of an axion in this model.
Indeed, we will find a modulated instability of the type described in the previous section.

The asymptotic behavior of the fields is given by
\bea
\psi(r) & \sim & \psi_{\infty}+mr^{\Delta_{+}}-c_{\psi}r^{\Delta_{-}} \\ 
z(r) &\sim& z_{0} +\frac{f_1 f_2}{r} \\
a_{0}(r) &\sim&  \mu -\frac{d}{r} \,,
\eea
where the boundary value $\psi_\infty$ and the exponents $\Delta_\pm$ are fixed by the 
fluxes $f_1,f_2$:
\bea
(f_1^2 + 4\cos^4\psi_\infty)\sin^2\psi_\infty =  (f_2^2 + 4\sin^4\psi_\infty)\cos^2\psi_\infty \label{const_solution} \\[5pt]
\Delta_\pm  =  -\frac{3}{2}\pm \frac{1}{2}\sqrt{9+16\frac{f_1^2
+16\cos^6\psi_\infty-12\cos^4\psi_\infty}{f_1^2+4\cos^6\psi_\infty}} \,. \label{deltapm}
\eea
The parameters $m$ and $c_\psi$ correspond to the ``mass" and ``condensate" of the fundamental fermions, respectively, and $\mu$ and $d$ to the 
chemical potential and charge density, respectively.\footnote{The physical charge density is given by
$d_{physical}=8\pi^{3} L^4 \alpha' T_{7} d$.}
Throughout the paper we will choose the fluxes $f_1, f_2$ such that $\Delta_+=-1$ (and $\Delta_-=-2$). 
We will vary $f_1$ and fix $f_2$ to satisfy this condition.

\subsection{Embeddings}

There are two kinds of embeddings in general:
``black hole" (BH) embeddings, in which the D7-brane crosses the horizon,
and ``Minkowski" (MN) embeddings, in which the D7-brane ends smoothly outside the horizon.
The latter exist only for $f_1=0$ or $f_2=0$ and for a fixed charge density to magnetic field ratio (filling fraction).
Here we will only consider BH embeddings with a vanishing magnetic field $b=0$.\footnote{The fluctuations of
the MN embeddings (at zero temperature) were studied in \cite{Jokela:2010nu}.}

For the fluctuation analysis, it is convenient to use the coordinate $u\equiv r_T/r$.
The boundary is now at $u=0$, and the horizon is at $u=1$.
The embedding equations of motion now read
\bea
 -u^2 r_T z'(u) \equiv \bar {z}'(u)
 &=& -\frac{f_1 f_2 h(u)}{\hat g(u)} \nonumber\\
-\frac{u^2}{r_T} a'_0(u) \equiv \bar{a}'_0(u)
&=& \hat d\, \frac{h(u)}{\hat g(u)}\nonumber\\
u^6\partial_u\left(\hat g(u) \psi'(u)\right) &=&  \frac{h(u)}{2\hat g(u)}\partial_\psi G(u) \ ,
\eea
where for convenience we have defined $\bar{a}'_{0}$ and $\bar{z}'$ and the prime now denotes differentiation with respect to $u$.
We have also defined $\hat{d}=\frac{d}{r_{T}^{2}}$.
The functions $G(u)$ and $\hat g(u)$ are defined slightly differently than in \cite{Jokela:2010nu}: 
\bea
G(u) &=& (f_1^2+4\cos^4\psi)(f_2^2+4\sin^4\psi)\nonumber\\ [5pt]
\hat g(u) 
&=& \frac{h}{u^2}\sqrt{ \frac{\hat{d}^2 u^4+G-f_1^2 f_2^2 h }{1+ h u^2\psi'^2}  } \ .
\eea

The presence of the horizon imposes the boundary condition
\be
\psi'(1) = -\frac{1}{8}\frac{\partial_\psi G(1)}{\hat{d}^2+G(1)} \,.
\ee
The mass associated with a given solution is given by
\be
\hat m \equiv m r_T^{\Delta_+} \equiv u^{\Delta_+}\left(\psi(u)-\psi_\infty\right)\Big|_{u\to 0} \ .
\ee

The BH embedding corresponds to the metallic phase of the model.
In particular, it has a longitudinal conductivity at zero magnetic field given by  \cite{Bergman:2010gm}
\be
\label{sigmaxx_BH}
\sigma_{xx} = \frac{N_3}{2\pi^2} r_T^{-2}
\sqrt{d^2 + r_T^4 G(1)}\,.
\ee
At low temperature and non-zero charge density 
\be
\sigma_{xx} 
\sim \frac{d}{T^2} \ ,
\ee
which is 
the correct behavior for a Fermi liquid.
Note, however, that this behavior is universal for any probe brane, and in particular holds also in cases where
the fundamental degrees of freedom include bosons \cite{Karch:2009eb}.
The heat capacity for a class of similar theories was computed in \cite{Karch:2008fa} and was found to behave at finite density and at low temperature as $C_{v} \sim \frac{T^4}{d}$, which is different from a regular Fermi liquid.

\section{Quasi-normal modes}
In this section we will study the fluctuations of all the fields around the BH embedding described in the previous section. 
Due to rotational invariance, we can restrict to fluctuations propagating only in the $x$-direction; we thus consider fluctuations of the form 
$f(u)e^{-i\omega t+ikx}$.
We rescale the fluctuations 
\be
\delta\hat z\equiv r_T\delta z, \ \ 
\delta\hat a_{t,x,y}\equiv \frac{\delta a_{t,x,y}}{r_T}, \ \ 
\delta\hat e_x\equiv \frac{\delta e_x}{r^2_T} \equiv \frac{k\delta a_t + \omega\delta a_x}{r_T^2},
\ee
and define
$\hat\omega \equiv \omega/ r_T$ and 
$\hat k \equiv k/r_T$.
This will remove $r_T$ from all the fluctuation equations.
It is worth recalling here the relationship to the physical momenta
$\omega_{phys}=\pi T \hat\omega$ and  $k_{phys}=\pi T \hat k$.
We also define the function 
\be
\hat A(u) \equiv 1+ h(u) u^2\psi'(u)^2 + h(u) u^{-4}\bar z'(u)^2-\bar a'_0(u)^2 \,.
\ee

\subsection{Fluctuation equations}

Expanding the action to second order in fluctuations, we obtain a set of coupled,
linearized equations for the fluctuations. 
The equation of motion for $\delta\psi$ is given by
\bea
& & \left(- \frac{h}{2\hat g u^4}\left(\partial^2_\psi G-\frac{1}{2G}(\partial_\psi G)^2\right)+\frac{u^2}{2}\partial_u\left[\hat g\psi'\partial_\psi\log G\right]\right)\delta\psi =  \nonumber\\
& &-u^2\partial_u\left[\frac{\hat g}{\hat A}\left(1+ h u^{-4}\bar z'^2-\bar a'^2_0\right)\delta\psi'\right] \nonumber\\
& & +\frac{\hat g u^2}{ h^2}\left[-\left(1+ hu^{-4}\bar z'^2\right)\hat\omega^2+\left(1+ hu^{-4}\bar z'^2-\bar a'^2_0\right) h \hat k^2\right]\delta\psi \nonumber\\
& &-\frac{\hat g}{2u^2}\partial_\psi\log G\bar z'\delta \hat z'+ \frac{\hat g\bar z'\psi'}{ h}\left[-\hat\omega^2
+ h\hat k^2\right]\delta\hat z -u^2\partial_u\left[\frac{\hat g h}{\hat A u^2}\psi' \bar z'\delta \hat z'\right] \nonumber\\
& &+\frac{\hat d u^2}{2}\partial_\psi\log G\delta\hat a'_t+u^2\partial_u\left[\frac{\hat d  h u^2}{\hat A} \psi'\delta\hat a'_t\right]-\hat k\frac{\hat d u^4}{2}\psi'\delta\hat e_x-i \hat k 4\bar a'_0\sin^2(2\psi)\delta\hat a_y \,.
\label{deltapsi}
\eea
The equation of motion for $\delta\hat z$ is given by
\bea
& &  -\frac{\hat g}{h}\bar z'\psi'\left(-\hat \omega^2+\hat h\hat k^2\right)\delta\psi+u^2\partial_u\left[\frac{\hat g h}{\hat A u^2}\bar z'\psi'\delta\psi'-\frac{\hat g}{2u^4}\partial_\psi\log G \bar z'\delta\psi\right] = \nonumber \\
& &
+\frac{\hat g}{h^2}\left(-\left(1+ hu^2\psi'^2\right)\hat\omega^2+\left(1+ hu^2\psi'^2-\bar a'^2_0\right) h \hat k^2\right)\delta\hat z \nonumber \\
& &-u^2\partial_u\left[\frac{1+ hu^2 \psi'^2-\bar a'^2_0}{\hat A}\hat g u^{-2}\delta \hat z'\right] +\hat k\hat d \bar z' \delta\hat e_x-u^2\partial_u\left[\frac{\hat d h}{\hat Au^2}\bar z'\delta\hat a'_t\right] \ .
\label{deltaz}
\eea
The equation of motion for $\delta\hat a_t$ is given by
\be
 u^2\partial_u\hat H 
+\hat d\hat k^2(\bar z'\delta\hat z-u^4\psi'\delta\psi)
 -\hat k\frac{\hat gu^4}{h^2}\left(1+ hu^{-4}\bar z'^2+ hu^2\psi'^2\right)\delta\hat e_x
 -i\hat ku^2\delta\hat a_y\partial_u\left[2c(u)\right] = 0 \,. 
 \label{deltaat}
 \ee
where
\be
 \hat H(u) 
 = \frac{\hat gu^2}{\hat A h}\left(1+ hu^{-4}\bar z'^2+ hu^2\psi'^2\right)\delta \hat a'_t -\frac{\hat d}{2}\partial_\psi\log G\delta\psi+\frac{\hat d h u^2}{\hat A}\psi'\delta\psi'-\frac{\hat d h}{\hat A u^2} \bar z'\delta\hat z' .
\ee
The equation of motion for $\delta\hat a_x$ is given by
\bea
 & &
\hat d\hat k\hat\omega(\bar z'\delta\hat z-u^4\psi'\delta\psi)
-\hat\omega\frac{\hat gu^4}{ h^2}\left(1+ hu^{-4}\bar z'^2+ hu^2\psi'^2\right)\delta\hat e_x \nonumber\\
 & &+\frac{u^2}{\hat\omega}\partial_u\left[\hat gu^2(-\delta\hat e'_x+\hat k\delta\hat a'_t)\right]-i\hat\omega u^2\delta\hat a_y\partial_u\left[2c(u) \right]=0.
 \label{deltaax}
 \eea
Finally, the equation of motion for $\delta\hat a_y$ is given by
\bea
 & & i\hat k 4\sin^2(2\psi)\bar a'_0\delta\psi+i\delta\hat e_xu^2\partial_u\left[2 c(u) \right]-\frac{\hat gu^4}{ h^2}\left(1+ hu^{-4}\bar z'^2+ hu^2\psi'^2\right)\hat\omega^2\delta\hat a_y  \nonumber\\
& & +\frac{\hat gu^4}{ h}\hat A\hat k^2\delta\hat a_y-u^2\partial_u\left[\hat gu^2\delta \hat a'_y\right] = 0\ .
 \label{deltaay}
\eea
There is also a constraint coming from the gauge choice $a_u=0$:
\be
-\hat\omega\hat H+\frac{\hat k}{\hat\omega}\hat gu^2\left(-\delta\hat e'_x+k\delta\hat a'_t\right) = 0 \ .
\ee

\subsection{Method of solution}

In this subsection we will briefly describe how the above equations of motion are solved to find quasi-normal modes. 
Although it is true that in some cases the
equations of motion decouple from one another, we do not separately discuss those special cases. 
The method that we will describe here works for any 
number of coupled or decoupled differential equations.

First of all, we are interested in normalizable solutions with  infalling boundary conditions
for the different fields. By inspection, one finds that all the fields, except for $\delta\hat a_t$,\footnote{Notice that the 
different fall-off guarantees that the temporal component of the gauge field vanishes at the horizon, as should be the case when the thermal circle pinches off.} will have the same infalling behavior as $u\to 1$:
\bea
 \delta\psi,\delta\hat z,\delta\hat e_x,\delta\hat a_y & \sim & (1-u)^{-i\frac{\hat\omega}{4}} \\
 \delta\hat a_t & \sim & (1-u)^{1-i\frac{\hat\omega}{4}} \ .
\eea
We separate out this leading singular behavior. For example,
$\delta\psi= (1-u)^{-i\frac{\hat\omega}{4}}\delta\psi_{reg}$, where 
$\delta\psi_{reg}$ is regular at the horizon.

We are interested in finding those solutions for which all the fluctuations have a normalizable behavior near the AdS boundary. 
Generically choosing
boundary conditions at the horizon and shooting towards $u\to 0$ would be hopeless. 
However, one can utilize the so-called determinant method \cite{Amado:2009ts,Kaminski:2009dh}, 
which was succesfully applied also in \cite{Jokela:2010nu}. 
Here we have five field equations for five fields 
($\delta\psi_{reg},\delta\hat z_{reg},\delta\hat e_{x,reg},\delta\hat a_{y,reg},\delta\hat a_{t,reg}$) to 
solve, but they are subject to a constraint. 
Imposing the constraint for this system is equivalent to imposing it on the boundary conditions at the horizon
and satisfying all the other equations of motion.
Thus, there are two alternative routes one can choose for taking the constraint into account: 1) make use of the 
constraint to solve for $\delta\hat a'_t$ in terms of all the other fields or 2) only impose the constraint on the horizon boundary conditions 
for the temporal gauge field. Both
of the routes are equivalent, but we found it numerically faster to follow the second path. Therefore, we will not lose any equations but make sure to take the
constraint into account. According to the determinant method, we choose a set of linearly independent boundary conditions at the horizon, say,
\be\label{eq:bc}
 \{\delta\psi_{reg},\delta\hat z_{reg},\delta\hat e_{x,reg},\delta\hat a_{y,reg}\} = \{(1,1,1,1),(1,1,1,-1),(1,1,-1,1),(1,-1,1,1)\} \ .
\ee
As mentioned above, the boundary condition for $\delta\hat a_{t,reg}$ at the horizon is set by the constraint: 
By expanding the constraint as a power series around the horizon, at zeroth 
order, one finds a condition for $\delta\hat a_{t,reg}(1)$ in terms of others 
(this condition is equivalent to the relationship one would find from the $\delta\hat a_t$ equation of motion (\ref{deltaat})).
The derivatives at the horizon are set by the equations of motion. 
For example, $\delta\psi'_{reg}(1)$ is found by expanding the $\delta\psi$ equation of motion (\ref{deltapsi})
around the horizon, and $\delta\hat a'_{t,reg}(1)$ is set by the boundary conditions and the derivatives of the other fields.

Finally, for any given momentum we solve the set of differential equations four times, corresponding to the four different boundary conditions in (\ref{eq:bc}). The
interesting object to look at is the determinant
\be
 \det \left.\left( \begin{array}{cccc}
 u^{\Delta_+}\delta\psi_{reg}^I & u^{\Delta_+}\delta\psi_{reg}^{II} & u^{\Delta_+}\delta\psi_{reg}^{III} & u^{\Delta_+}\delta\psi_{reg}^{IV} \\
 \delta\hat z_{reg}^I & \delta\hat z_{reg}^{II} & \delta\hat z_{reg}^{III} & \delta\hat z_{reg}^{IV} \\ 
 \delta\hat e_{x,reg}^I & \delta\hat e_{x,reg}^{II} & \delta\hat e_{x,reg}^{III} & \delta\hat e_{x,reg}^{IV} \\
 \delta\hat a_{y,reg}^I & \delta\hat a_{y,reg}^{II} & \delta\hat a_{y,reg}^{III} & \delta\hat a_{y,reg}^{IV}
\end{array} \right)\right|_{u\to 0} 
\ee
at the AdS boundary. For a given $\hat k$, one then begins to scan over the complex valued energy $\hat\omega$ until a zero of the determinant is found. 
Once this is the case, one concludes that a normalizable solution has been found.
There is a linear combination of the boundary conditions giving the desired normalizable solution,
for which all fluctuations vanish at the AdS boundary.

In practice, we start with a limit of the parameters such that the equations decouple
and consider separately the different fluctuations. The quasi-normal modes 
are identified as the values of $(\hat{\omega},\hat{k})$ where the contribution to the determinant changes sign.
The accuracy of these positions is therefore determined by the resolution of the scan,
which in our case is at least $10^{-3}$.
Away from this limit the determinant is complex in general, and we look for positions at which
its absolute value is smaller than some ``numerical zero". The ``numerical zero" is chosen so as
to maintain the same accuracy as in the decoupled limit.

\subsection{Hydrodynamical mode and zero sound}

Generically, the equations for the fluctuations are all coupled. 
However, there are certain limits of the parameters $(\hat d,\hat m,\hat k,f_1)$ where some of the equations decouple 
from the rest. This enables us to associate, even in the range of parameters where the equations are coupled, 
certain modes to particular fields at least by name. But the reader should not get confused; when the system is 
in the range of parameters where the equations are coupled, this naming may be meaningless.

\begin{figure}[ht]
\center
\includegraphics[width=0.40\textwidth]{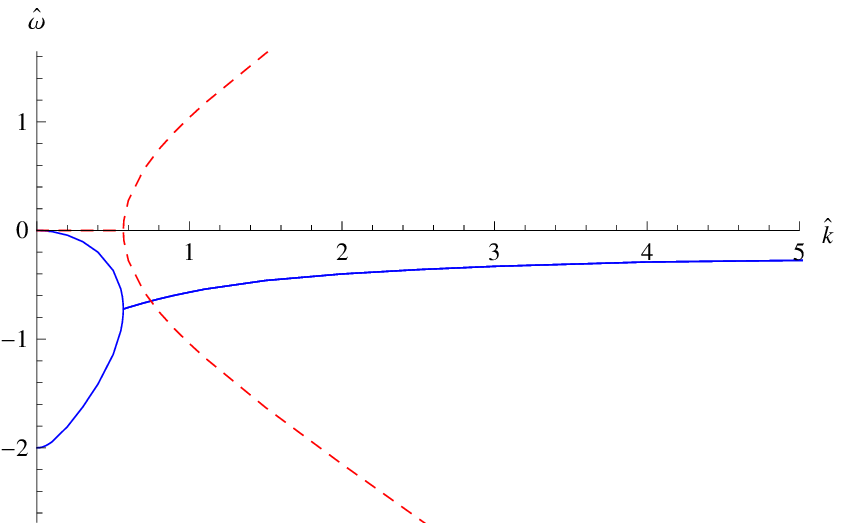}
\includegraphics[width=0.40\textwidth]{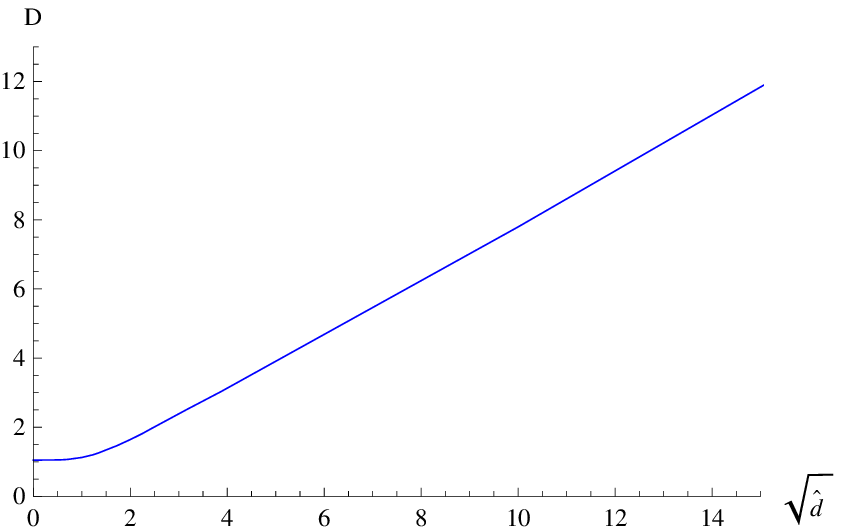}
\caption{Left: Quasi-normal modes for $\delta\hat e_x$ at $\hat{d}=0$, $\hat{m}=0$ and $\psi_{\infty}=\pi/4$. 
Blue (solid) lines are the imaginary parts and red (dashed) lines are the 
real parts. Right: Diffusion constant as a function of $\sqrt{\hat d}$. 
Notice the linear behavior for small temperatures, $\hat d\gg 1$.}
\label{hydro}
\end{figure}

Let us start with the case where $\hat m=0$ and $\hat d=0$. For this particular case $\psi'=0$ and $\bar a'_0=0$, so the equation 
for $\delta\hat e_x$ decouples
from the rest. For $\delta\hat e_x$ we find a hydrodynamical mode (\emph{i.e.}, $\hat \omega(\hat k\to 0)\to 0$) associated with charge conservation, 
which is purely imaginary and behaves as
\begin{equation}
 \hat\omega=-i D\hat k^2+\ldots \;,\;\;\; \hat k\sim 0 \ ,
\end{equation}
where $D$ is the diffusion constant.  For low $\hat k$, this is the important mode 
since it has the lowest imaginary value, corresponding to the collective mode with the longest lifetime. However, as $\hat k$ is increased 
this mode meets another mode which initially started at  $\hat \omega (\hat k=0)\sim -2 i$.
After the modes meet they become a pair of complex modes with both real and imaginary parts; the real parts have the same magnitude but opposite signs. 
We depict this in Fig.~\ref{hydro} (left panel).

\begin{figure}[ht]
\center
\includegraphics[width=0.40\textwidth]{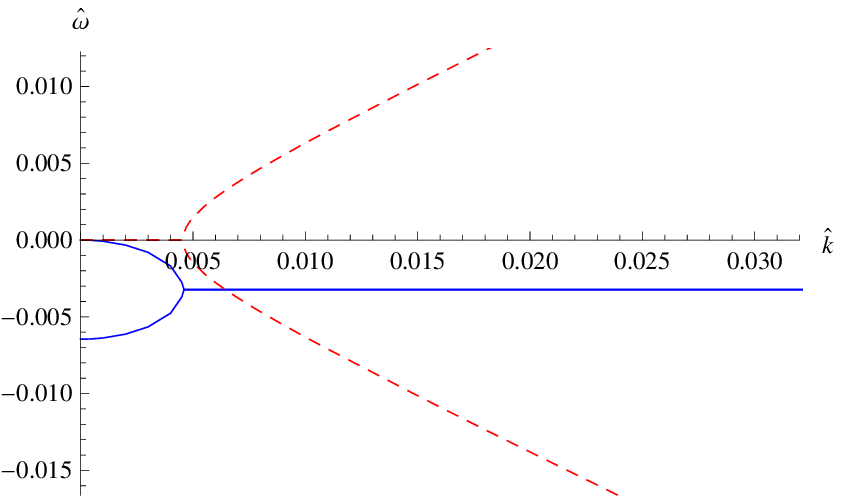}
\includegraphics[width=0.40\textwidth]{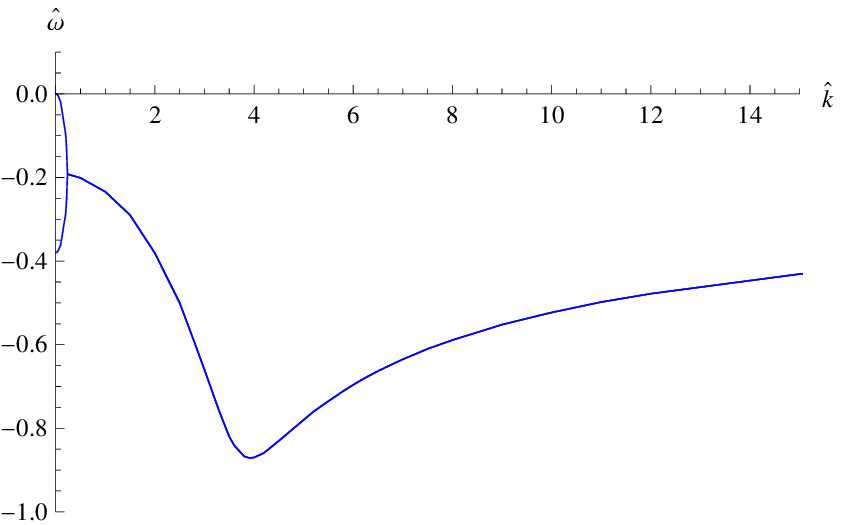}
\caption{Left: Quasi-normal mode for the coupled ($\delta\hat e_x,\delta\hat z)$ system at  $\hat m=0$, $\hat d=10^4$ and $\psi_{\infty}=\pi/4$.  
Right: Imaginary part of the quasi-normal 
mode of the $\delta\hat e_x,\delta\hat z$ system, with $\hat m=0$,  $\hat{d}=5$ and $\psi_{\infty}=\pi/4$.
}
\label{zero}
\end{figure}

When $\hat d \neq 0$, but still $\hat m=0$, the longitudinal vector and the $\delta\hat z$ scalar are coupled but the quasi-normal mode 
structure (if $\hat\omega$ is not too large) is quite similar to the zero-density case. There is a purely imaginary 
hydrodynamical mode that meets at some finite $\hat k$ another purely imaginary, but non-hydrodynamical, mode, and for larger $\hat k$ 
there are two complex modes. This transition from purely imaginary modes to complex modes can also be regarded as a transition from 
a hydrodynamical regime to a collisionless regime (see \cite{Kaminski:2009dh} for a similar phenomenon in a (3+1)-dimensional boundary theory).  
The values of both $|\hat\omega|$  and $\hat k$ 
at the meeting point decrease with increasing $\hat d$; see the left panel of Fig.~\ref{zero}.

The imaginary part of the complex mode for $\hat{d}\neq 0$ is shown on the right of Fig.~\ref{zero}. 
The imaginary part decreases as $\hat{k}$ increases and 
then increases for very large $\hat{k}$ towards zero. As $\hat{d}$ is increased, the minimum of $\rm{Im} \ \hat\omega$  
shifts to larger $\hat{k}$.

Indeed, this mode has the properties of the holographic zero sound mode \cite{Karch:2008fa, kp}. 
As $T\to 0$ at fixed charge $d$ (and therefore 
$\hat{d}\rightarrow\infty$), the purely diffusive mode disappears completely and the complex 
pair of modes start at $\hat k \rightarrow 0$, with an approximate dispersion relation
\be
\hat{\omega}= v \hat{k}-i a \hat{k}^2+\ldots \ .
\ee
A calculation of $v$, for all $f_1$, gives $v\sim \frac{1}{\sqrt{2}}$ in the large $\hat{d}$ limit, in agreement with \cite{Karch:2008fa}. 
In this model, in addition to the quantum attenuation at zero temperature, there is very large damping for low momentum at non-zero temperature, turning the zero sound mode to a purely dissipative mode.

However, as we will see in the next subsection, these results can only be trusted up to some critical value of $\hat d$,
beyond which there is an instability towards a striped phase.

 \subsection{Stripes}

In this section we show that at large enough $\hat d$, an instability occurs. This instability occurs only at some 
non-zero $k_{phys}$,  similar to what was found in \cite{Nakamura:2009tf} but now in a (2+1)-dimensional theory. 
This indicates that the Fermi liquid is unstable and that the true ground state of the (2+1)-dimensional theory will be a 
striped phase, similar to a charge density and spin wave.
\begin{figure}[ht]
\center
\includegraphics[width=0.40\textwidth]{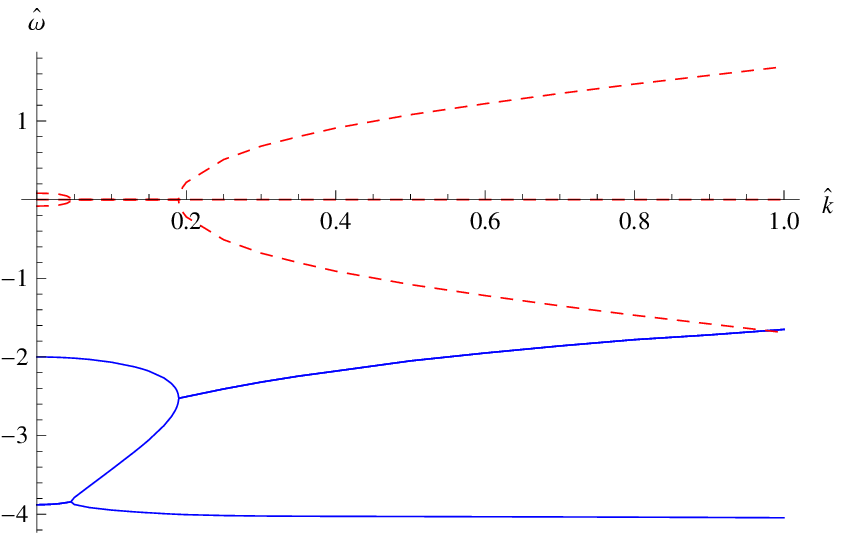}
\includegraphics[width=0.40\textwidth]{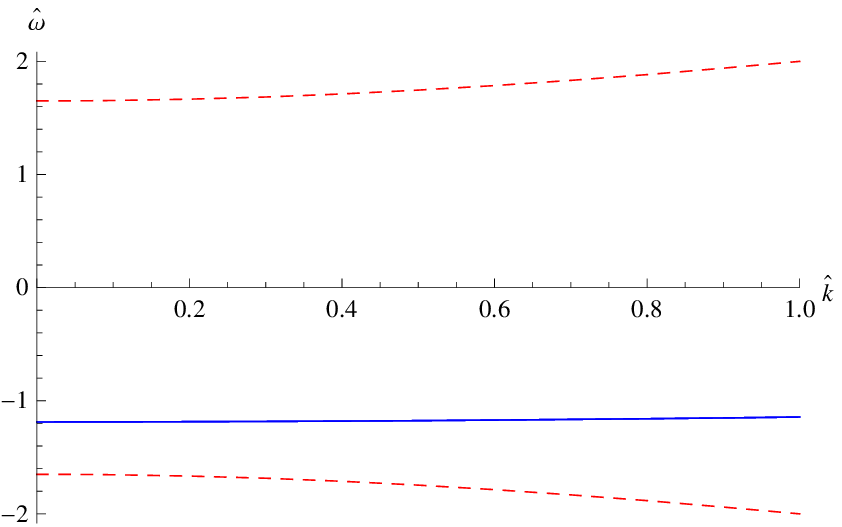}
\caption{Lowest quasi-normal mode 
for $\delta\hat a_y$ (left) and $\delta\psi$ (right), with $\hat d=0, \hat m=0$ and $\psi_{\infty}=\pi/4$.}
\label{aypsi}
\end{figure}

We start with $\hat d=0$ and $\hat m=0$. In this case the transverse gauge field fluctuation $\delta\hat a_y$ and the 
embedding scalar fluctuation $\delta\psi$ decouple. The lowest quasi-normal modes are portrayed in Fig.~\ref{aypsi}.
At $\hat k=0$ the lowest quasi-normal mode has the same value of $\hat\omega$ as the second quasi-normal mode of the longitudinal vector 
(equations (\ref{deltaax}) and (\ref{deltaay}) are the same in this limit).

\begin{figure}[ht]
\center
\includegraphics[width=0.40\textwidth]{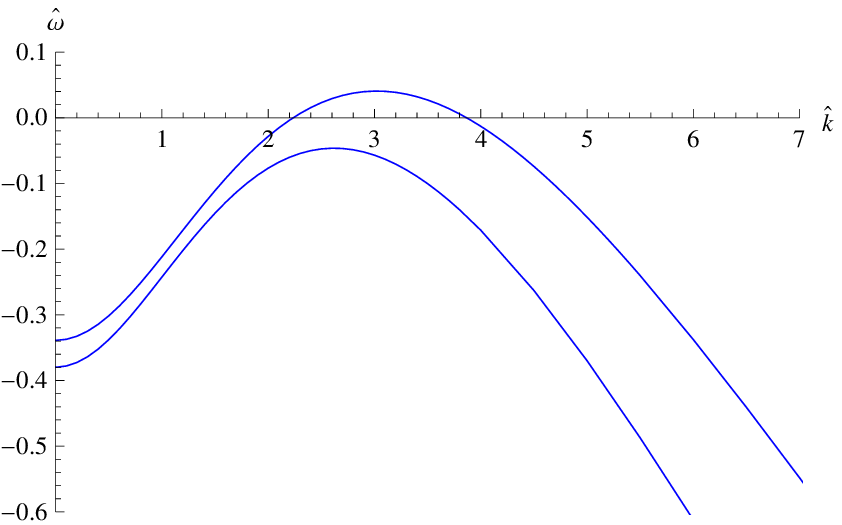}
\includegraphics[width=0.40\textwidth]{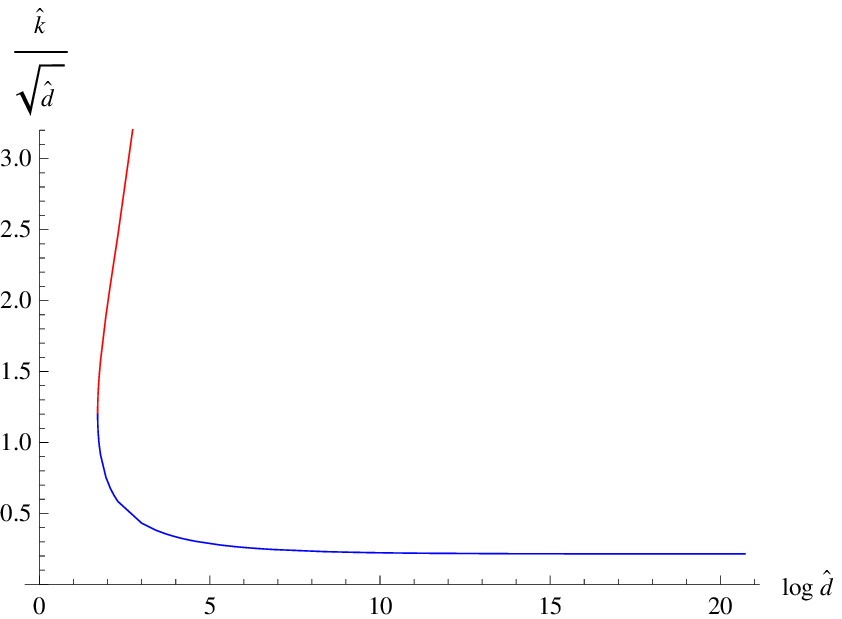}
\caption{Properties of the unstable mode with $\hat m=0$ and $\psi_{\infty}=\pi/4$.  Left: Lowest purely imaginary mode for the 
coupled $(\delta\hat a_y,\delta\hat\psi)$ system.  The lowest curve is for $\hat d=5$, and the upper 
curve is for $\hat d=6$. Right: Range of instability; $\hat k_{min}$ (blue) and $\hat k_{max}$ (red) are plotted against $\hat d$.} 
\label{ins}
\end{figure}

For $\hat d>0$ and $\hat m=0$ the modes $\delta\hat a_y$ and $\delta\hat\psi$ couple to one another. Above a critical value of $\hat d$,
the lowest negative imaginary quasi-normal mode crosses onto the upper half-plane, at some value of $\hat k$, 
and becomes a positive imaginary frequency mode, \emph{i.e.}, an instability (Fig.~\ref{ins}, left panel).
The 
critical value is $\hat{d}_c \simeq 5.5$. There is a range in $\hat k$ for which the 
frequency is positive imaginary, and at some larger $\hat k$ it crosses again back onto the lower part of the 
complex frequency plane. 
The dependence of this range 
on $\hat d$ is shown in Fig.~\ref{ins} (right panel).
In this range the system is driven towards a striped phase, where both the transverse current and the fermion
bi-linear are spatially modulated.
The latter, in particular, describes a spin density wave.
This can be seen, for example, from the fact that at finite density a spatially modulated field $\psi$ couples
linearly to the magnetic field through the bulk CS term in (\ref{CS_action}).

\begin{figure}[ht]
\center
\includegraphics[width=0.50\textwidth]{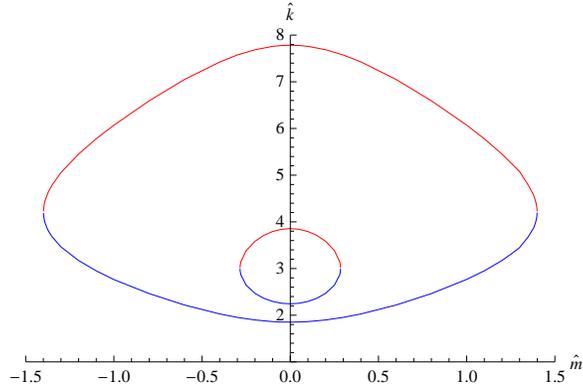}
\caption{Domains of instability for non-zero mass $\hat{m}$ and $\psi_\infty = \pi/4$ 
for $\hat{d}=6$ (smaller domain) and $\hat{d}=10$ (larger domain).} \label{kminkmax}
\end{figure}

The situation is similar when $\hat m\neq 0$. In this case all the fluctuations are coupled, and we still find an instability. 
As $|\hat m|$ increases, the critical value of $\hat d$ 
grows; see Fig.~\ref{kminkmax}. Since now $\delta \hat a_t$ is also involved, one also obtains a charge density wave.
As $T\to 0$ the instability appears for any small, non-zero charge density. The physical momentum at which this happens is given by 
$k_{phys} \simeq \frac{ 0.215 \sqrt{d}}{L} $. Things are qualitatively similar when $\psi_{\infty} \neq \pi/4$. For instance when one
decreases $\psi_\infty$ from the value $\pi/4$ then $\hat d_c$ decreases slightly.

\section{Conclusion}

The ungapped (BH embedding) phase of the D3-D7' system has an instability towards a striped phase
above a critical value of the ratio of the charge density to the temperature squared. 
For a non-zero mass the instability involves all the fields, and therefore one expects that the new phase will be such 
that all the physical quantities (normalizable modes) associated with these fields are spatially modulated. These include charge, spin, and the transverse currents.
In the dual bulk theory the instability 
is the result of an axion-like term in the action. 
On the other hand, the existence of this term is intimately related to the fermionic nature of the boundary theory.
We believe that striped phases are a general property of such theories.
Since the model should have a striped phase, one should be able to construct it as a solution of the equation of motion and study its properties.

\bigskip
\noindent {\bf \large Acknowledgments}

We thank Daniel Podolsky for useful comments and discussions.  
O.B. is supported in part by the Israel Science Foundation under grant no.~392/09,
and the US-Israel Binational Science Foundation under grant no.~2008-072.
N.J. has been supported in part by the Israel Science Foundation under grant no.~392/09 
and in part at the Technion by a fellowship from the Lady Davis Foundation.  
The work of G.L.  is supported in part by the Israel Science Foundation under grant no.~392/09.
The research of M.L. is supported by the European Union grant FP7-REGPOT-2008-1-CreteHEPCosmo-228644.


\begin{thebibliography}{99}

\bibitem{Hartnoll:2009sz}
  S.~A.~Hartnoll,
  Class.\ Quant.\ Grav.\  {\bf 26}, 224002 (2009).
  [arXiv:0903.3246 [hep-th]].
\bibitem{Liu:2009dm}
  H.~Liu, J.~McGreevy, D.~Vegh,
  Phys.\ Rev.\  {\bf D83}, 065029 (2011).
  [arXiv:0903.2477 [hep-th]].

\bibitem{Cubrovic:2009ye}
  M.~Cubrovic, J.~Zaanen, K.~Schalm,
  Science {\bf 325}, 439-444 (2009).
  [arXiv:0904.1993 [hep-th]].

\bibitem{D'Hoker:2010rz}
  E.~D'Hoker, P.~Kraus,
  JHEP {\bf 1005}, 083 (2010).
  [arXiv:1003.1302 [hep-th]].
\bibitem{Lifschytz:2009sz}
  G.~Lifschytz, M.~Lippert,
  Phys.\ Rev.\  {\bf D80 } (2009)  066007.
  [arXiv:0906.3892 [hep-th]].



\bibitem{Esko}  
  E.~Keski-Vakkuri and P.~Kraus,
  JHEP {\bf 0809}, 130 (2008)
  [arXiv:0805.4643 [hep-th]].

\bibitem{Davis}
  J.~L.~Davis, P.~Kraus, A.~Shah,
  JHEP {\bf 0811}, 020 (2008).
  [arXiv:0809.1876 [hep-th]].
  
  
\bibitem{Fujita}  
  M.~Fujita, W.~Li, S.~Ryu and T.~Takayanagi,
  JHEP {\bf 0906}, 066 (2009)
  [arXiv:0901.0924 [hep-th]].


\bibitem{Hikida}  
  Y.~Hikida, W.~Li and T.~Takayanagi,
  JHEP {\bf 0907}, 065 (2009)
  [arXiv:0903.2194 [hep-th]];
  
\bibitem{Alanen}  
 J.~Alanen, E.~Keski-Vakkuri, P.~Kraus {\it et al.},
JHEP {\bf 0911}, 014 (2009)
[arXiv:0905.4538 [hep-th]].


\bibitem{Bergman:2010gm}
  O.~Bergman, N.~Jokela, G.~Lifschytz and M.~Lippert,
  JHEP {\bf 1010} (2010) 063
  [arXiv:1003.4965 [hep-th]].
  
\bibitem{Jokela:2010nu}
  N.~Jokela, G.~Lifschytz and M.~Lippert,
  JHEP {\bf 1102} (2011) 104
  [arXiv:1012.1230 [hep-th]].
  
  
\bibitem{Jokela}  
  N.~Jokela, M.~J\"arvinen, M.~Lippert,
  JHEP {\bf 1105}, 101 (2011).
  [arXiv:1101.3329 [hep-th]].
  
\bibitem{Rey:2008zz}
  S.~J.~Rey,
  Talk at Strings 2007;
  Prog.\ Theor.\ Phys.\ Suppl.\  {\bf 177}, 128 (2009)
  [arXiv:0911.5295 [hep-th]].







\bibitem{Gruner:1994zz}
  G.~Gruner,
  Rev.\ Mod.\ Phys.\  {\bf 66}, 1-24 (1994).
\bibitem{Gruner:1988zz}
  G.~Gruner,
  Rev.\ Mod.\ Phys.\  {\bf 60}, 1129-1181 (1988).


\bibitem{Nakamura:2009tf}
  S.~Nakamura, H.~Ooguri, C.~-S.~Park,
  Phys.\ Rev.\  {\bf D81}, 044018 (2010).
  [arXiv:0911.0679 [hep-th]].


\bibitem{Ooguri:2010xs}
  H.~Ooguri, C.~-S.~Park,
  Phys.\ Rev.\ Lett.\  {\bf 106}, 061601 (2011).
  [arXiv:1011.4144 [hep-th]].
\bibitem{Ooguri:2010kt}
  H.~Ooguri, C.~-S.~Park,
  Phys.\ Rev.\  {\bf D82}, 126001 (2010).
  [arXiv:1007.3737 [hep-th]].
\bibitem{Bayona:2011ab}
  C.~A.~B.~Bayona, K.~Peeters, M.~Zamaklar,
  [arXiv:1104.2291 [hep-th]].


\bibitem{Donos:2011bh}
  A.~Donos and J.~P.~Gauntlett,
  arXiv:1106.2004 [hep-th].


  
  
  
\bibitem{Karch:2009eb}
  A.~Karch, M.~Kulaxizi, A.~Parnachev,
  JHEP {\bf 0911}, 017 (2009).
  [arXiv:0908.3493 [hep-th]].

\bibitem{Karch:2008fa}
  A.~Karch, D.~T.~Son, A.~O.~Starinets,
  [arXiv:0806.3796 [hep-th]];


\bibitem{Amado:2009ts}
  I.~Amado, M.~Kaminski and K.~Landsteiner,
  JHEP {\bf 0905} (2009) 021
  [arXiv:0903.2209 [hep-th]].
  
\bibitem{Kaminski:2009dh}
  M.~Kaminski, K.~Landsteiner, J.~Mas {\it et al.},
  JHEP {\bf 1002}, 021 (2010).
  [arXiv:0911.3610 [hep-th]].

\bibitem{kp}
  M.~Kulaxizi and A.~Parnachev,
  Phys.\ Rev.\  D {\bf 78} (2008) 086004
  [arXiv:0808.3953 [hep-th]].

\end{thebibliography}
\end{document}